# Creation of nano-skyrmion lattice in Fe/Ir(111) system using voltage pulse

Yulong Yang[1], Mingming Shuai[1], Haiming Huang[1], Rui Song[1], Yi Zhu[1], Yanghui Liao[1], Yinyan Zhu[1,2,4], Xiaodong Zhou[1,2,4*], Lifeng Yin[1,2,3,4,5*] and Jian Shen[1,2,3,4,5*]

[1] State Key Laboratory of Surface Physics, Institute for Nanoelectronic Devices and Quantum Computing, and Department of Physics, Fudan University, Shanghai 200433, China.
[2] Shanghai Qi Zhi Institute, Shanghai 200232, China
[3] Shanghai Research Center for Quantum Sciences, Shanghai 201315, China.
[4] Zhangjiang Fudan International Innovation Center, Fudan University, Shanghai 201210, China.
[5] Collaborative Innovation Center of Advanced Microstructures, Nanjing 210093, China.

*Emails: zhouxd@fudan.edu.cn, lifengyin@fudan.edu.cn, shenj5494@fudan.edu.cn

## Abstract

Magnetic ultrathin films grown on heavy metal substrates often exhibit rich spin structures due to the competition between various magnetic interactions such as Heisenberg exchange, Dzyaloshinskii–Moriya interaction and higher-order spin interactions. Here we employ spin-polarized scanning tunneling microscopy to study magnetic nano-skyrmion phase in Fe monolayer grown on Ir(111) substrate. Our observations show that the formation of nano-skyrmion lattice in the Fe/Ir(111) system depends sensitively on the growth conditions and various non-skyrmion spin states can be formed. Remarkably, the application of voltage pulses between the tip and the sample can trigger a non-skyrmion to skyrmion phase transition. The fact that nano-skyrmions can be created using voltage pulse indicates that the balance between the competing magnetic interactions can be affected by an external electric field, which is highly useful to design skyrmion-based spintronic devices with low energy consumption.

Magnetic ultrathin films grown on heavy metal substrates have been a fruitful playground to study magnetism in reduced dimensions [1,2]. The magnetic ground state of such seemingly simple system turns out to be complex as it is determined by various magnetic interactions, such as Heisenberg exchange, Dzyaloshinskii–Moriya interaction (DMI) and even higher order spin interactions. If the Heisenberg exchange dominates, collinear ferromagnetic (FM) or antiferromagnetic (AFM) states are formed. When the DMI becomes comparable to the exchange interaction, various non-collinear spin states can form [3-6]. The competition between various magnetic interactions is pronounced in Fe/Ir(111) ultrathin film system, where the hybridization between the 3d state of Fe and the 5d state of Ir suppresses the Heisenberg exchange interaction in the Fe layer enabling spin interactions of higher order to play a crucial role in determining the magnetic ground state of the system. It has been experimentally observed that Fe islands with monolayer thickness are in a magnetic nano-skyrmion state [7-12], and Fe islands with bi-layer or tri-layer thickness are in a spin-spiral state [13,14]. A recent experiment further indicates that the formation of the nano-skyrmion state is strongly influenced by the Fe adatom diffusion rate during the growth [15]. The nano-skyrmion state only exists in Fe monolayer islands with a perfect triangular shape. In other words, different non-skyrmion spin states may exist in Fe monolayer islands on Ir(111) substrate that compete with the nano-skyrmion state.

Magnetic skyrmion can be utilized as the information bits in future spintronic devices, for which an electrical manipulation of skyrmion (creation, annihilation and movement) is essential for its integration into modern electronic technology [16-22]. An early study of Pd/Fe bilayer on Ir(111) showed that individual skyrmion can be written and deleted in a controlled way by local spin-polarized currents from a scanning tunneling microscopy (STM) tip [23]. Considering the high energy dissipation and the Joule heating in such an electric current involved operation, a pure electric field method is called for to achieve the low energy consumption. This was later realized in another STM study, in which an individual magnetic skyrmion in a tri-layer Fe island on Ir(111) can be switched reversibly from the skyrmion state to the FM state by a tip induced local electric field [14]. These experiments demonstrate that the Fe/Ir(111) system is a fertile playground for exploring the electrical manipulation of skyrmion. It is noted that the skyrmion phase in the aforementioned works only exist under an external magnetic field. In this regard, the Fe monolayer on Ir(111) is unique as it exhibits a skyrmion phase at zero field which is also beneficial for the application. Therefore, it would be interesting to investigate the Fe monolayer on Ir(111) and see if its skyrmion phase can be electronically manipulated.

In this work, we report the use of a voltage pulse from a STM tip to transit a non-skyrmion metastable spin state to the magnetic nano-skrymion state in Fe monolayer islands on Ir(111) substrate. Once formed, the skyrmion phase remains unchanged upon further applications of voltage pulses, indicating that the skyrmion state is likely a true ground state in Fe monolayer islands on Ir(111).

Fe was thermally deposited with a nominal thickness of sub-monolayer on an Ir(111) single crystal surface. Prior to the Fe deposition, the Ir(111) single crystal was prepared by cycles of $Ar^+$ ion sputtering and annealing at $T = 1500$ K. It was further annealed at oxygen atmosphere ($2 \times 10^{-6}$ mbar) at a slightly lower temperature (1200~1400 K) to get rid of the surface impurities, which is crucial to obtain the skyrmion lattice [15]. Fe was evaporated from a rod heated by an e-beam bombardment. The deposition was conducted when the substrate was at room temperature. A bulk Fe tip with 0.25 mm in diameter was used in the STM measurement, whose magnetic moment at the tip apex points along its axis resulting in a sensitivity to the out-of-plane component of the sample magnetization. This spin-polarized STM (SP-STM) enables us to resolve the magnetic structure in real space at the atomic scale [24]. All SP-STM experiments were performed at 5 K in the absence of any external magnetic field.

Figure 1(a) shows a SP-STM constant-current image of typical Fe islands grown on an Ir(111) substrate. Two triangular shaped monolayer islands point in opposite directions, corresponding to face centered cubic (fcc) and hexagonal closed packed (hcp) stackings, respectively. An island formed by second layer Fe adatoms is observed in the middle of the fcc island. A differentiated image of Fig. 1(a) is displayed in Fig. 1(b) for a better visualization of the signal contrast. In the fcc island, a square lattice is observed in the monolayer region, which was known as magnetic nano-skyrmion state [7-9]. Its magnetic origin is confirmed by a field-dependent measurement (see supplementary S1). This nano-skyrmion lattice has multiple domains with different orientations (denoted by the green arrows in Fig. 1(b)) because its four-fold symmetry is different from the three-fold symmetry of the underlying hexagonal atomic lattice of Fe/Ir(111). The hcp island displays another nano-skrymion state with a hexagonal lattice [25]. The nano-skyrmion phase is totally absent in the bi-layer region where reconstruction lines along three different orientations are observed. It is a spin spiral state according to previous studies [13,14].

In addition to the aforementioned results consistent with previous reports, we have new observations in the Fe monolayer islands on Ir(111). Figure 1(c) shows a SP-STM constant-current image of a different fcc Fe monolayer island. Compared to Fig. 1(a), we modify the color scheme of Fig. 1(c) to enhance the relatively weak signal contrast on the island. A similar color scheme is adopted in other SP-STM constant-current images of this paper whenever applicable. Figure 1(c) displays a spatial coexistence of "bright" and "dark" regions without the regular nano-skyrmion lattice. Instead, one sees a small labyrinthine pattern in the "dark" region, and a bubble pattern in the "bright" region. To simplify the data description and discussions hereafter, we refer the states without a clear skyrmion lattice as a non-skyrmion phase. Strikingly, we have observed that the non-skyrmion phase can be transited to the skyrmion phase by a tip voltage pulse across the STM tip-sample tunneling junction. Figure 1(d) show a SP-STM constant-current image of the same Fe island after the application of a +6 V voltage pulse (50ms duration, sample is positive while the tip is grounded) to the lower-right corner of the island (red circle in Fig. 1d). A regular skyrmion lattice is formed in regions surrounding the voltage pulse location. The rest areas closer to the upper-left edge of the island remains in the non-skyrmion phase. A second voltage pulse is then applied to upper-right corner of the island as shown in Fig. 1(e). After the second voltage pulse, the island rearranges its spin states by rotating the pattern anti-clockwise from the lower-right corner to the upper-right corner. The skyrmion lattice also "moves" to the upper-right corner accordingly. While the non-skyrmion to skyrmion phase transition occurs in regions near the voltage pulse location under the +6 V voltage pulse, the whole island can be converted into a clean and uniform skyrmion phase by applying a higher voltage pulse (+8 V) to its center (Fig. 1f). Interestingly, this tip voltage pulse induced skyrmion lattice only displays a single domain with one orientation, different from the multi-domain state which is often observed in the as-grown Fe islands (e.g. Fig. 1a-b). This is a clear indication that the tip voltage pulse induced skyrmion lattice is formed by nucleation and growth of a single skyrmion domain while an as-grown Fe island may host skyrmion domains nucleated from different regions of the island.

The tip voltage pulse induced non-skyrmion to skyrmion phase transition occurs not only in fcc Fe islands, but also in hcp Fe islands and in Fe strips grown along the Ir(111) substrate step edge. Figure 2 shows a large sample area across an Ir(111) substrate step edge before (Fig. 2a-c) and after (Fig. 2d-f) multiple tip voltage pulses (+8 V, 50 ms duration). Both fcc and hcp Fe monolayer islands exist, together with a Fe strip along the Ir(111) step edge. While a clear skyrmion lattice cannot be observed in the as-grown state of these Fe areas (Fig. 2a-b), it appears after the tip voltage pulse (Fig. 2d-e) in all Fe islands and strip. Such a skyrmion lattice differs in their electronic structure between fcc and hcp Fe areas, giving rise to a signal contrast in the dI/dU map (Fig. 2f).

Figure 3 shows the detailed processes of the non-skyrmion to skyrmion phase transition upon a series of voltages pulses applied to the same location of an Fe island with an increasing amplitude from -1 V to -7 V. This Fe island was originally in a non-skyrmion state with a mixture of the "bright" and the "dark" regions. The tip voltage pulse applied to the left corner of the island (red circle in Fig. 3b) changes the spatial arrangement of the electronic states, i.e., the "bright" and electronically more uniform phase is aggregated to the left corner. The boundary between the "bright" and the "dark" regions gradually shifts to right as the voltage increases, which can be considered as an evidence of transition from the "dark" phase to the "bright" phase. The exact nature of the spin states of these phases are not known at this stage. At -7 V, a skyrmion lattice starts to appear within the "bright" region. We note that a similar phase evolution happens at the positive tip voltage pulses (see supplementary S2).

The tip voltage pulse induced non-skyrmion to skyrmion phase transition is irreversible. As shown in Fig. 4, a tip voltage pulse is applied to transform a non-skyrmion phase (Fig. 4a) to a clean and uniform skyrmion phase (Fig. 4b). Once formed, the skyrmion phase remains unchanged upon further applications of voltage pulses (Fig. 4c). This can be understood by considering the free energy profile as sketched at the bottom of Fig. 4(a-c). As mentioned above, the delicate balance between different competing magnetic interactions in Fe/Ir(111) system determines its magnetic ground state. The system could be easily trapped in many metastable spin states with close energy levels such as the "bright" and the "dark" non-skyrmion phases. The skyrmion phase is the true magnetic ground state of the system with the lowest free energy. However, an energy barrier exists between them preventing the system from reaching its ground state. The applied voltage pulse supplies the required energy to overcome the barrier and drive the system into its ground state. The fact that the opposite transition from a skyrmion to non-skyrmion phase cannot happen under the same tip voltage pulse indicates that this ground state is very stable. Actually, such a skyrmion phase in Fe monolayer on Ir(111) can sustain under an external field of up to +9 T [2].

Finally, we discuss the kinetic nature of the tip voltage pulse induced non-skyrmion to skyrmion phase transition. The tip voltage pulse yields both a large current density (on the order of $10^2$ A/cm$^2$) and a large electric field (on the order of 0.5 V/Å). Both the current and the electric field can drive the non-skyrmion to the skyrmion phase transition. In the former case (i.e., current), the spin-transfer torque (STT) from the spin-polarized current is often discussed [26]. However, the fact that the phase transition occurs under both positive and negative voltage pulses in a very similar manner implies that STT is unlikely responsible for the phase transition because STT depends on the spin polarization of the tunnel current and its direction. A local temperature increase caused by the high current density (Joule heating) can play a role too. However, with the long pulse duration (50 ms) one expects the whole island reaches the same temperature and a spatially uniform skyrmion phase should be formed after the transition, which is not the case in our experiment (Fig. 1d-e). In the latter case (i.e., electric field), the electric field can cause a structural relaxation of Fe atom as reported in Fe/Cu(111) system[27]. This can affect the magnetic exchange interactions. The DMI can also be affected because it stems from the inversion symmetry breaking at the buried Fe/Ir interface (Rashba DMI) and is therefore sensitive to the vertical electric field applied by the tip. However, such electric field will decay strongly with the distance from the surface due to the screening effect from the metallic Fe layer. The change of DMI is thus very limited. Since we do observe a change of the apparent height in SP-STM constant-current image during the phase transition in our Fe/Ir(111) system, we speculate a similar mechanism, i.e., an electric field induced structural relaxation of Fe atom, to be responsible for the non-skyrmion to skyrmion phase transition. Such an intimate interplay between the structural and the magnetic degree of freedom in the formation of skyrmion phase in the Fe/Ir(111) system has been revealed before, e.g., the different skyrmion lattice symmetry between fcc and hcp Fe areas on Ir(111) [25].

In summary, we have observed various spin states in the as-grown Fe monolayer islands on Ir(111). A tip voltage pulse can be used to trigger a non-skyrmion to skyrmion phase transition. Such a transition is irreversible indicating skyrmion as a true magnetic ground state in the system. It is argued that, this transition happens as a result of an electric field induced structural relaxation of Fe atom which affects the competing magnetic interactions. This finding adds another evidence of the feasibility of manipulating magnetic skyrmions on the nanometer scale via an electric manner.

**Acknowledgement**

This work was supported by National Key Research Program of China (2022YFA1403300), the National Natural Science Foundation of China (11427902, 11991060, 12074075, 12074080 and 12274088), the Shanghai Municipal Science and

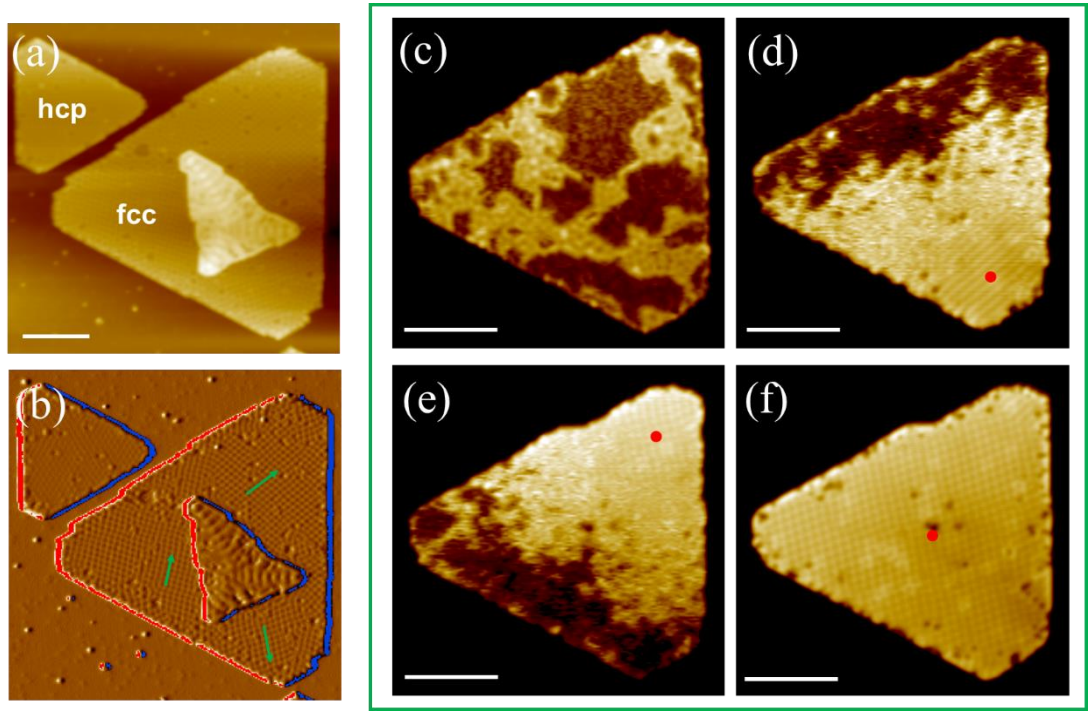

FIG. 1. (a-b) SP-STM constant-current image and the corresponding differentiated image of Fe monolayer and double layer grown on an Ir(111) substrate ($U=50$ mV, $I=100$ pA). (c-f) SP-STM constant-current images of an fcc stacking Fe monolayer island subjected to different tip voltage pulses. $U=50$ mV, $I=100$ pA for (c) and $U=100$ mV, $I=200$ pA for (d-f). The scale bar is 10 nm.

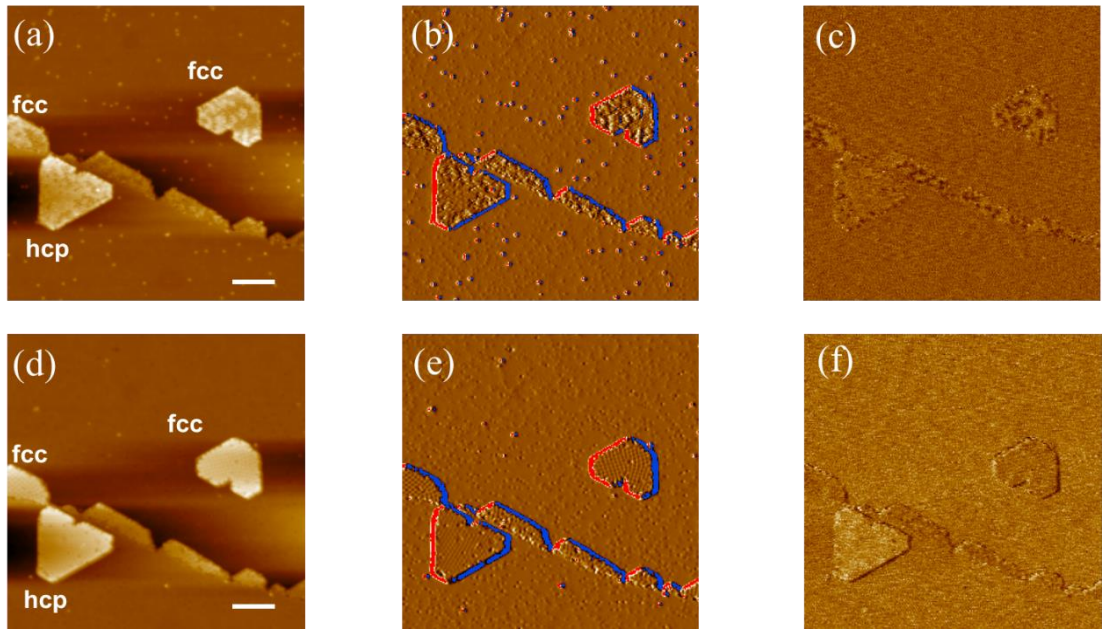

FIG. 2. (a-c) SP-STM constant-current image, the differentiated image and the dI/dU map of a large sample area before the application of a tip voltage pulse. (d-f) SP-STM constant-current image, the differentiated image and the dI/dU map of a large sample area after the application of a tip voltage pulse. *U=50 mV, I=200 pA* for all and the scale bar is 10 nm.

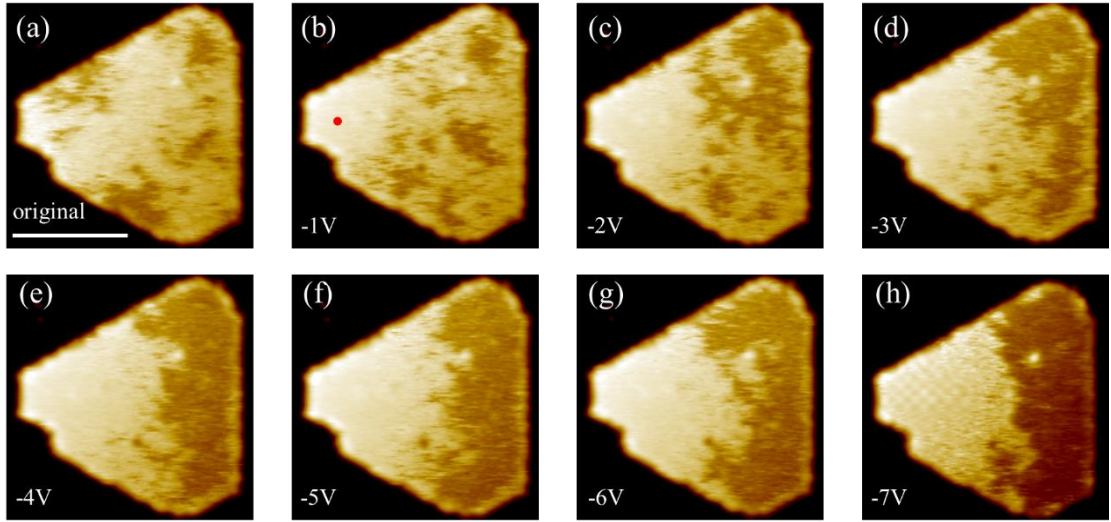

FIG. 3. (a-h) A series of SP-STM constant-current images of an Fe island under different tip voltage pulses with an increasing amplitude as denoted in the lower-left corner of the image. *U=100 mV, I=200 pA* for all and the scale bar is 10 nm.

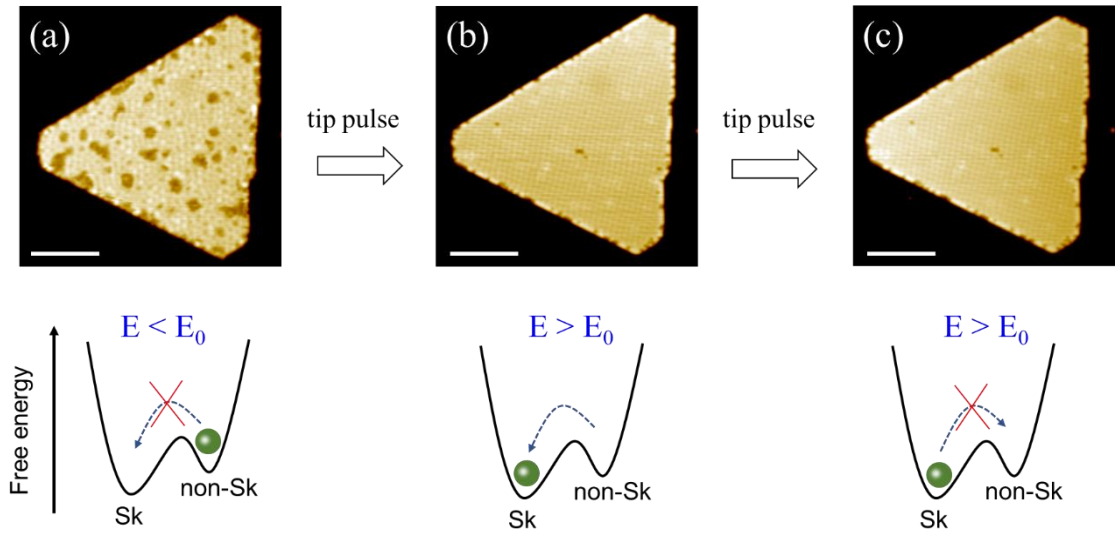

FIG. 4. (a-c) SP-STM constant-current images of an Fe island under two consecutive tip voltage pulses. The schematic of the free energy profile of different spin states is shown at the bottom. *U=50 mV, I=100 pA* for all and the scale bar is 10 nm.